\documentclass{PoS}

\newcommand{\gr}{$\gamma$-ray \,}

\title{Cosmic-ray driven winds}

\ShortTitle{Cosmic-ray driven winds}

\author{\speaker{Heinrich V\"olk}%
         \thanks{I would like to thank Vladimir Zirakashvili for discussions on
           the gas and cosmic-ray halo of the Milky Way.}\\
        Max-Planck-Institut fuer Kernphysik\\
        E-mail: \email{Heinrich.Voelk@mpi-hd.mpg.de}}

      \abstract{The theory of Galactic Winds, driven by the cosmic-ray pressure
        gradient, is reviewed both on the magnetohydrodynamic and on the
        kinetic level. In this picture the magnetic field of the Galaxy above
        the dense gas disk is assumed to have a flux tube geometry, the flux
        tubes rising locally perpendicular out of the disk to become radially
        directed at large distances, with the cosmic-ray sources located deep
        within the Galactic disk. At least above the gas disk, the magnetic
        fluctuations which resonantly scatter the cosmic rays are
        selfconsistently excited as Alf{`e}n waves by the escaping cosmic
        rays. The fluctuation amplitudes remain finite through nonlinear wave
        dissipation. The spatially increasing speed of the resulting outflow
        results in a diffusion-convection boundary whose position depends on
        particle momentum. It replaces the escape boundary of static diffusion
        models. New effects like overall Galactic mass and angular momentum
        loss as well as gas heating beyond the disk appear. Also particle
        re-acceleration in the distant wind halo suggests itself. The resulting
        magnetohydrodynamic flow properties and the cosmic-ray transport
        properties are compared with observations. On the whole they show
        remarkable agreement. General limitations and generalisations of the
        basic model arise due to the expected simultaneous infall of matter
        from the environment of the Galaxy. On an intergalactic scale the
        combined winds from the Local Group galaxies should form a ``Local
        Group Bubble``. Its properties remain to be studied in detail.}

\FullConference{Cosmic Rays and the InterStellar Medium\\
                 24-27 June 2014\\
                 Montpellier, France}

\begin{document}

\section{Characteristics of the Galactic cosmic rays}
The Galactic cosmic rays (CRs) represent a flux of relativistic, fully ionized
atomic nuclei with a power-law type energy distribution, impinging on the
Earth's atmosphere. The flux of relativistic electrons is about 1 \% of this by
number. The total CR energy density is $E_\mathrm{c}\sim 1 \, \mathrm{eV}
\rm{cm}^{-3} \sim B^2/8\pi \sim E_{\mathrm{th}} \sim E_{\mathrm{turb}} \sim
E_{\mathrm{rad}}$ in the neighbourhood of the Solar System. $\gamma$-ray
observations with satellite detectors at high energies (HE:$E_{\gamma} <
100$~GeV) suggest that this is true $\sim$ everywhere in the diffuse
Interstellar Medium (ISM) of the Galactic gas disk, outside of the sources.

The momentum distribution is essentially {\it isotropic}, presumably due to
frequent directional scattering in the irregular magnetic field, anchored in
the {\it thermal} ionized gas.  This implies a diffusive propagation process
relative to the scattering centers of the ISM and eventual advection with these
scattering centers. As a result the CRs constitute a nonthermal, relativistic
gas of high pressure which is equal partner in the dynamics of the ISM.

The chemical abundances around 1 GeV/n are similar to those of the Solar System
and of the ISM which implies that the sources mainly accelerate ordinary
interstellar matter. However, there are notable exceptions: in particular the
light CR nuclei Li, Be, B are strongly overabundant relative to this cosmic
abundance. They are therefore interpreted as secondary spallation products of
heavier primary CR nuclei, predominantly in the interstellar gas. The so-called
grammage encountered by those particles at energies of a few GeV/n is $\simeq 8
\mbox{g cm}^{-2}$ which is small compared to the spallation mean free
path. Thus for the majority of particles there is only limited confinement in
the Galactic gas disk of density $\simeq 1~\mathrm{particle \, cm}^{-3}$, where
they spend a few million years. The abundance ratio of secondary to primary CR
nuclei, like the B/C-ratio, decreases with particle energy above about 1 GeV:
higher-energy particles leave the spallation region faster than lower-energy
particles and therefore the source spectra must contain more high-energy
particles. Recent indications are that this does not continue into the TeV
range \cite{Derbina05,Obermeier11,Aguilar13}, which is likely due to the fact
that secondary particles produced {\it inside} the sources are also
re-accelerated there to a resulting hard spectrum,
e.g. \cite{Berezhko03,Blasi09,Berezhko14}. An empirical conclusion for the CR
source energy spectrum is then: $dN/dE \propto (E/n)^{-\gamma}$, with $\gamma
\simeq 2.1 \, \mathrm{to} \, 2.2$, at least up to energies $E/n \sim
\mathrm{few} \times 10^2$~GeV.

The absolute life time of the particles that return to the Solar System at a
few $100 \mathrm{MeV/n}$ is approximately $1.5 \times 10^7$~yrs from an
analysis of the fraction of the radioactive ${^{10}}\mbox{Be}$ nuclei in the
arriving CRs, see e.g. \cite{Yanasak01}. This suggests that we have to imagine
that even nonrelativistic CRs predominantly propagate in a low-density halo
before ``escaping''. For higher-energy particles this appears to be even more
likely.

\section{CR propagation and the Galactic Wind}

The traditional picture holds that CRs propagate diffusively from the sources
in the dense gas disk into a fixed, static confinement volume. From there they
escape to infinity upon reaching the ``boundary''of this halo. This picture is
kinematically consistent. For a recent review, see \cite{Strong07}. 

The alternative picture, which we shall discuss here, is a dynamic one of a
nonlinear character: the gas and CR pressure gradients jointly drive a Galactic
Wind (GW) in which the CRs are diffusively confined by {\it self-excited}
magnetic field fluctuations, the relativistic CRs trying to establish an
infinite scale height.

A basic question is then, whether the CRs escape with the gas, producing ever
more extending magnetic loops as a result of the Parker instability
\cite{Parker66}, or whether they escape by carrying only a very small amount of
gas with them, effectively in a boyant bubble of reconnected magnetic field,
leaving behind a dynamo for the Galactic magnetic field. Presumably both CR
loss processes operate in the Galaxy with comparable CR removal rates, as
argued in \cite{Breitschwerdt93}.


The first analytical attempt to describe a CR-driven wind flow was made in
spherical symmetry \cite{Ipavich75}. Later on wind flows were studied in a
Galactic disk symmetry (including a dark matter halo), assuming flow tube
configurations out of the disk that allowed a local description. Flow tubes
start perpendicular to the disk in z-direction with constant cross-section
$A(s)$ along flow lines $s$ to become spherical, $A(s) \propto s^2$ at
distances exceeding the disk radius $\simeq 15$~kpc
\cite{Breitschwerdt91}. These solutions were subsequently generalized to
include Galactic rotation and the corresponding azimuthal magnetic field
effects, as well as gas heating due to wave dissipation \cite{Zirakashvili96},
and CR kinetics \cite{Ptuskin97}.


 \begin{figure}[thb]
  \centering \includegraphics[width=9cm]{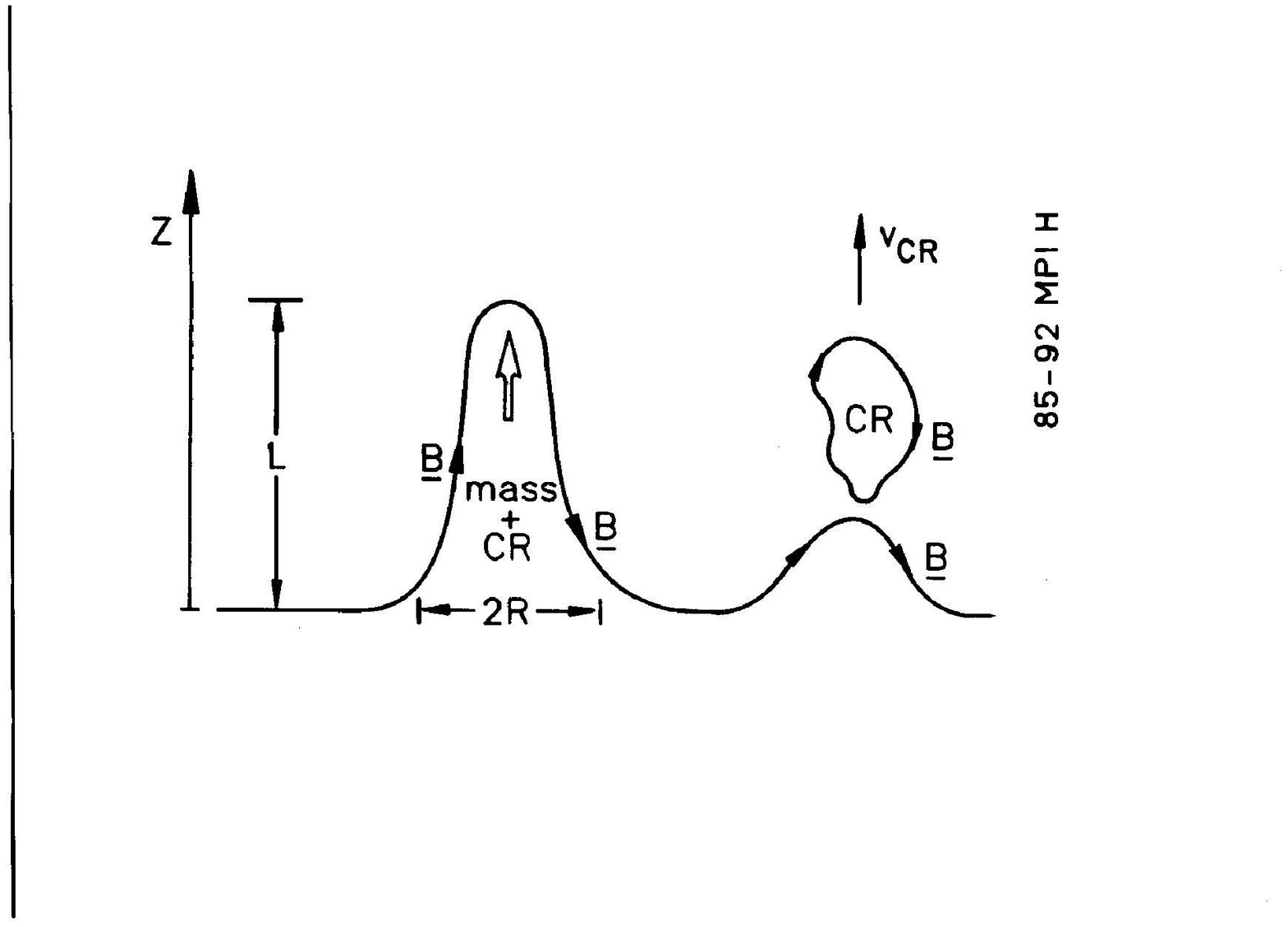}
  \centering \includegraphics[width=5cm]{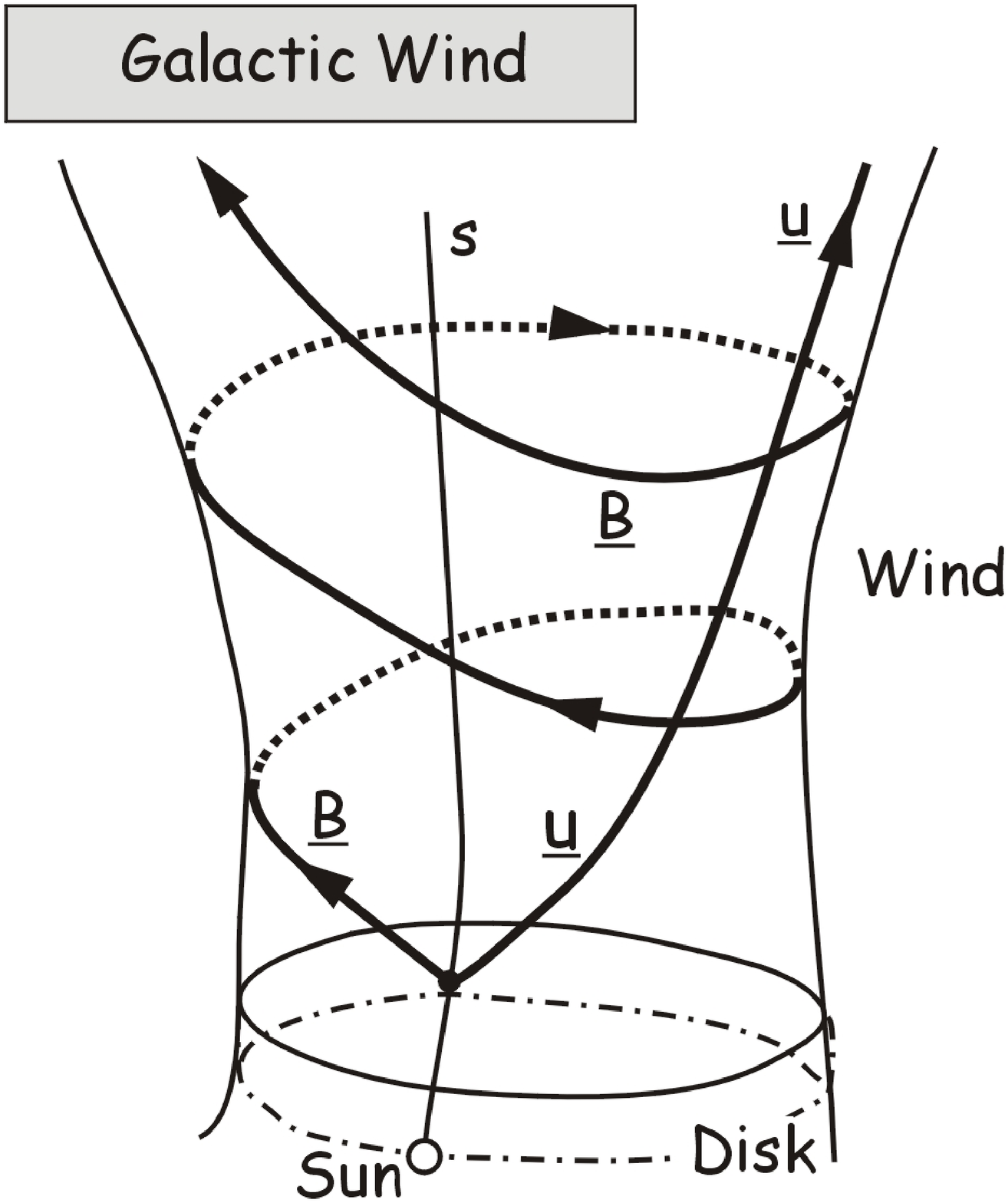}
  \caption{Left: Extension of magnetic field lobes in z-direction,
    perpendicular to the Galactic disk. In the left part of the figure the
    lobes are expanding ultimately to infinity with the CRs and the thermal gas
    in a Galactic Wind. In the right part the competing magnetic reconnection
    process allows CRs to move out by boyancy without a substantial
    accompagnying gas component (from \cite{Breitschwerdt93}).
  Right: Schematic of a CR-driven Galactic Wind from the Sun in the inner
    disk. The coordinate along flow lines is $s$, with the flow velocity
    $\underline{u}$ increasing with $s$, while the magnetic field
    $\underline{B}$ develops into a so-called Parker spiral. The flow is
    assumed to become radial at large $s$, starting perpendicular to the disk
    at low heights. (Adapted from \cite{Ptuskin97}).}
  \label{fig1}
\end{figure}

The flow is mainly driven by the outward force of the CR pressure gradient
$-\mathrm{grad} \, P_\mathrm{c}$. Outward diffusion of the CRs excites
resonantly scattering Alfv\'enic field fluctations in a selfconsistent way: $u
\cdot \mathrm{grad} \, B^2/8\pi = - v_\mathrm{A} \cdot \mathrm{grad} \,
P_\mathrm{c}$ in the halo plasma above the dense gas disk; here $v_\mathrm{A}$
denotes the Alfv\'en speed. Eventually the mass overburden is small enough to
be lifted up in the form of a wind. The amplitude of the waves remains finite
due to nonlinear wave dissipation, especially {\it nonlinear Landau damping} of
the Alfv\'en waves \cite{Lee73,Ptuskin97} which keeps the thermal gas warm/hot
despite the adiabatic expansion. Ultimately, for distances $s > 20$~kpc, the
outflow becomes supersonic. The total mass loss rate from the Galaxy is
estimated to be $\simeq 1 \mbox{M}_{\odot}$/yr, while the asymptotic wind
velocity $u(\infty) = \mbox{few} \, 100 \, \mbox{km/s}$ is of the order of the
escape velocity. The angular momentum los is about 20 \% of the Galactic total
in a Hubble time for the present-day parameters of the Galaxy
\cite{Zirakashvili96}. The wind termination shock is estimated to be roughly at
a distance of $300$ kpc. The wind velocity decreases over the disk with
increasing Galactrocentric radius. Therefore, the density of CRs in the Galaxy
should be rather independent of radius, despite the enhanced star formation
rate in the inner Galaxy. This is also a possible explanation for the weak
Galactocentric gradient of the diffuse \gr emission \cite{Breitschwerdt02}.

\begin{figure}
 \includegraphics[width=1\textwidth]{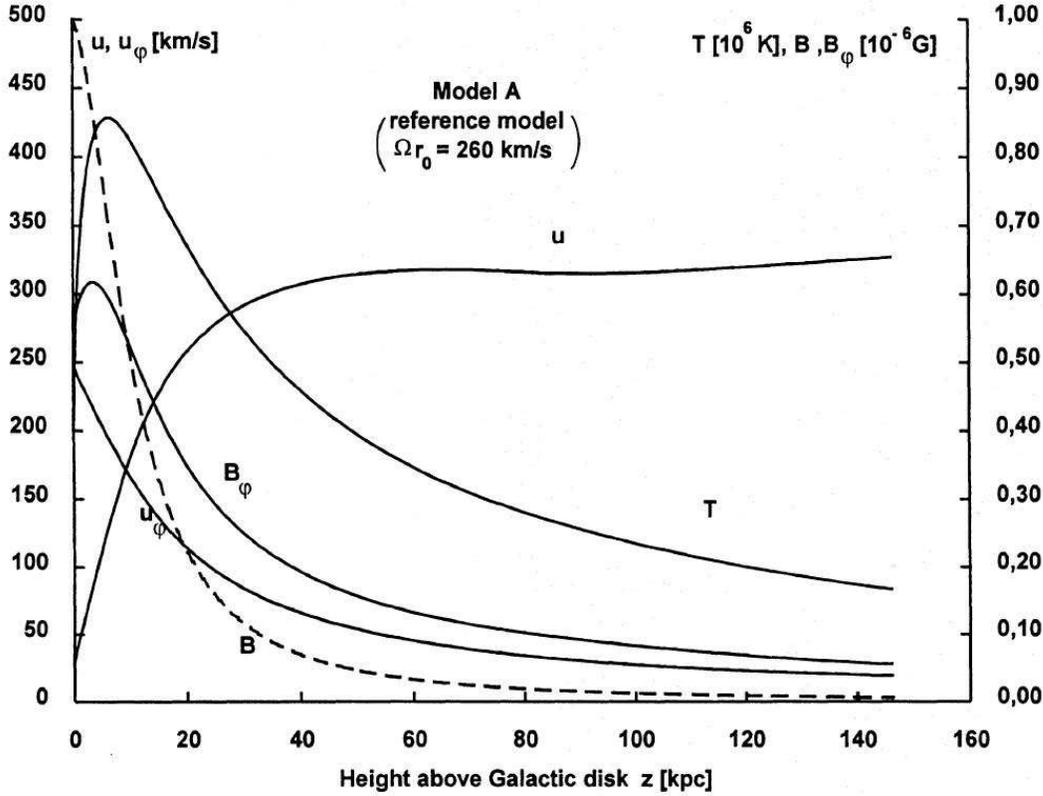}
 \caption{Variation of the meridional and azimuthal flow velocities $u$ and
   $u_{\varphi}$, azimuthal and meridional magnetic field strengths
   $B_{\varphi}$ and $B$, and gas temperature $T$, with distance $z$ from the
   the disk, for a rotational velocity of the Sun $\Omega r_0 = 260$~km/s at a
   galactocentric distance $r_0 = 8.5$~kpc \cite{Zirakashvili96}.}
\label{fig2}
\end{figure}

Globally speaking, the resulting CR transport is diffusive close to the
Galactic disk. It becomes convective further out, i.e. without return of
particles to the disk. The characteristic spatial boundary $s^{\star}$ between
these two propagation modes occurs where the diffusion time equals the
convection time and is roughly given by $(s^{\star})^2/\kappa \sim
s^{\star}/u(s^{\star})$. Here $\kappa(p)$ is the selfconsistent diffusion
coefficient along the flow lines $s$. It depends on particle momentum $p$. For
the momentum spectrum $Q(p) \propto p^{- \gamma}$ of the CR sources deep in the
disk, $s^{\star} \propto p^{(\gamma-3)/2)}$ and $\gamma \simeq (2
\gamma_{\rm{d}} + 3) / 3 \simeq 4.1$, where $\gamma_{\mathrm{d}} \simeq 4.7$ is
the observed spectral index at the Solar System. This means that the
diffusion-advection boundary moves out with increasing energy, approximately
$s^{\star} \propto p^{0.55}$ as long as the flow velocity increases \~ linearly
with $s$ /, ; at 1 TeV $s^{\star} \simeq 15$~kpc. However, the diffusion time
to reach the boundary $s^{\star}(p)$ is approximately energy-independent
$\simeq 2 \times 10^7$~yr and is roughly consistent with the ${^{10}}\mbox{Be}$
survival fraction at 100 MeV/n. In this way a natural boundary condition for
the propagation of the CRs in the Galaxy appears. It goes together with a
general mass and angular momentum loss phenomenon for a normal star forming
galaxy like the Milky Way, even at the present epoch of its evolution.

The local increase of the CR intensity above the rotating spiral arms in the
disk leads to a series of shocks in the wind that re-accelerate CRs produced in
these regions of enhanced star formation \cite{Voelk04}. Similarly, time
variations of the conditions at the base of the wind due to assumed localised
starbursts should induce strong wind shocks that also accelerate particles
beyond the ``knee'' of the observed spectrum \cite{Dorfi12}.

\subsection{The real world of the Galactic environment}

Obviously such models are necessarily idealized, in spite of their comparably
sophisticated physics. Reality should be more complex. For example the local
picture shows so-called high-velocity clouds falling back to the disk,
presumably through radiative cooling of hot upward expanding material which
gives rise to ``Galactic fountains''. In addition, a population of ``very
high-velocity clouds'' is observed which may represent true accretion of the
Galaxy, perhaps from the Magellanic Stream. {\it Infall} in this or other forms
appears also required to slow down the chemical evolution of the Galaxy, in
particular to keep the D/H-ratio high in the ISM \cite{Geiss02}. In addition, a
diffuse, hot Galactic gas halo has been inferred to exist with a wind-like
density profile $\propto r^{-2.1}$ and a total mass within $\sim 200$~kpc of
$\leq 10^{10} \mbox{M}_{\odot}$ \cite{Miller13}. It may be the result of a
Galactic mass loss rate of $\simeq 1 \mbox{M}_{\odot}$/yr on average over a
Hubble time, but might also be a reservoir of ongoing accretion. The observed,
diffuse soft X-ray and radio continuum emissions also require a global halo
model. Longitude-averaged GW models have been constructed to explain the
latitude profiles of the ROSAT $0.65$ keV and $0.85$ keV bands together with
the $408$ MHz radio band \cite{Everett09}. Such fits appear to require a
concentration of the wind to disk radii $R$ within rather narrow limits $3.5 <
R < 4.5$~kpc, with the B-field assumed to be vertical and decreasing $\propto
z^{-2}$. It will be interesting to see how such restricted base regions for the
GW can be made consistent with the known radial dependence of the star
formation conditions in the disk.

\section{A Local Group Bubble}

The distance of the Galactic Wind termination shock lies at several $100$ kpc
-- about half way to M31. Beyond this a shocked wind bubble develops. Therefore
it is likely that the wind bubbles of the Local Group galaxies push against
each other, so to say shoulder to shoulder. Possibly some winds even interact
directly. In any case, the wind bubbles should merge into a single ``Local
Group Bubble'' and the nuclear CRs may play a dynamical role in this
structure. The radio synchrotron emission is rather uncertain due to
synchrotron and Inverse Compton losses over these large distances, despite the
expected occurrence of re-acceleration processes in the winds and their
termination shocks. We should therefore look forward to the prospect of
determining the {\it low-frequency} Extragalactic radio synchrotron morphology
on scales of the Local Group of galaxies. LOFAR and SKA should be very
suitable instruments for this endeavour.

\end{document}